\documentclass[showpacs,amssymb,amsmath,nobibnotes, aps,prl,secnumarabic,superscriptaddress
,twocolumn%
]{revtex4}

\usepackage{bm}
\usepackage{natbib}
\usepackage{graphicx}

\begin{document}
\textfloatsep 10pt

\title{Turbulence-condensate interaction in two dimensions}
\author{H. Xia}
\author{H. Punzmann}
\affiliation{
Physical Sciences and Engineering, Australian National
University, Canberra ACT 0200, Australia}
\author{G. Falkovich}
\affiliation{Physics of Complex Systems, Weizmann Institute of
Science,  Rehovot 76100, Israel}
\author{M.G. Shats}
\affiliation{
Physical Sciences and Engineering, Australian National
University, Canberra ACT 0200, Australia}

\date{\today}

\begin{abstract}
We present experimental results on turbulence generated in thin fluid layers in the presence of a large-scale coherent flow, or a spectral condensate. It is shown that the condensate modifies the third-order velocity moment in a much wider interval of scales than the second one. The modification may include the change of sign of the third moment in the inverse cascade. This observation may help resolve a controversy on the energy flux in mesoscale atmospheric turbulence (10-500 km): to recover a correct energy flux from the third velocity moment one needs first to subtract the coherent flow. We find that the condensate also increases the velocity flatness.
\end{abstract}


\maketitle

More often than not, turbulence coexists with a flow coherent across the system size.
Such flows can be externally generated or appear as a result of spectral condensation due to an inverse turbulent cascade (see e.g. \cite{Hossain83,Sommeria86,Falkovich92,Smith_Yakhot_94,Chertkov07,Shats07}). Understanding interaction between turbulence and a mean flow is of prime importance for many problems in astrophysics, geophysics, plasma confinement, etc. From a fundamental viewpoint, it is interesting to understand how spectral condensation breaks symmetries that emerge in inverse cascades \cite{BBCF}, and how the condensate suppresses turbulence level and turbulent fluctuations \cite{DF,Terry_00,Shats07}. 
From a practical viewpoint, atmospheric physics presents arguably the most important cases of turbulence which interacts with the large-scale flows.  

Atmospheric motions are powered by gradients of solar heating. Vertical gradients cause thermal convection on the scale of the troposphere depth (less than 10 km). Horizontal gradients excite
motions on the scales from 1000 to 10000 km. Both inputs
are redistributed over wide spectral intervals by nonlinear
interactions \cite{Kraichnan67,Lilly83,Gage_Nastrom86}. The
wavenumber spectra measured in the upper troposphere and in the
lower stratosphere have shown two power laws: $E(k) \propto
k^{-5/3}$ for the scales between 10 and 500 km, and a steeper
spectrum with $E(k) \propto k^{-3}$ in the range
(500-3000) km (similar to the spectra in Fig.2) \cite{NG84}. Interestingly, such spectra appear in
the Kraichnan theory of 2D turbulence \cite{Kraichnan67}, where
$E(k)=C\epsilon ^{2/3}k^{-5/3}$ corresponds to an inverse energy
cascade and $E(k)=C_q \eta ^{2/3}k^{-3}$ to a direct vorticity
cascade, $\epsilon$ and $\eta$  being the dissipation rates of
energy and enstrophy respectively. This prompted a two-source
picture of atmospheric turbulence with a planetary-scale source of
vorticity and depth-scale source of energy, where the
large-scale spectrum is due to a direct vorticity
cascade \cite{Lilly89}. Alternatively, that spectrum can result from an inverse cascade of
inertio-gravity waves \cite{Falkovich92}. Yet another possibility is that the main energy at large scales is 
actually not in turbulent fluctuations but in a long-correlated flow (condensate), which can be either generated by external forces or appear in the process of spectral
condensation (turbulent counterpart of Bose-Einstein condensation), as suggested in \cite{Smith_Yakhot_94} and demonstrated experimentally here.
No less controversial is the nature of the mesoscale 5/3-spectrum. Is it an energy
cascade and what is the flux direction?

In homogeneous turbulence, spectral energy flux is expressed via
the third-order moment of the velocity \cite{Lindborg99}: $\epsilon = S_3/r$, where $S_3= \left [ \left (\langle (\delta V_L)^3 \rangle + \langle \delta
V_L (\delta V_T)^2 \rangle \right ) \right ]/2$. Here $\delta V_L$ and $\delta V_T$ are respectively longitudinal ($_L$) and transverse ($_T$) components of the velocity difference between points separated by $r$. Angular brackets denote time averaging. Positive
$S_3$ corresponds to the inverse energy cascade (from small to
large scales). Measurements of $S_3$ in the atmosphere gave a negative value in the interval 10-100 km, which was interpreted as the signature of the forward
energy cascade \cite{Cho_Lindborg01}. Here we demonstrate experimentally that a negative small-scale $S_3$ in a system with an inverse cascade can be caused by a large-scale shear flow.

Let us first consider how small- and large-scale parts of the velocity difference (respectively $\delta v$ and 
$\delta V$) contribute to the second and third velocity moments. The large-scale flow is spatially smooth
so that  $\delta V\simeq sr$ where $s=V/L_s$ is a large-scale velocity gradient and $L_s$ is the shear scale which depends on the system size and on the topology of the flow. Comparing $\langle(\delta V)^2 \rangle \cong s^2r^2$ with $\langle(\delta v)^2 \rangle \cong C(\epsilon r)^{2/3}$ we see that the small-scale (turbulent) part dominates at the scales smaller than $l_t \cong C^{3/4}s^{-3/2} \epsilon ^{1/2}$. For the third moment, we compare $\langle(\delta v)^3 \rangle \cong \epsilon r$ with the cross-correlation term $\langle \delta V(\delta v)^2 \rangle \cong srC(\epsilon r)^{2/3}$ and observe that the influence of $\delta V$ extends to a much smaller scale $l_* \cong C^{-3/2} s^{-3/2} \epsilon ^{1/2}$, because the dimensionless constant $C$ is substantially larger than unity, as discussed below. At the scale $l_*$, $\langle (\delta V)^3\rangle\simeq (sr)^3\ll \epsilon r$. That attests to a non-locality of condensate-turbulence interaction: not only condensate breaks scale-invariance of turbulence, but it imposes different scales on different moments.

The above estimates are true for a large-scale part produced by any source. In particular, when it is produced  by an inverse cascade (as in the experiments described below) one estimates $s$ as follows. Let the linear damping rate $\alpha$  be smaller than the inverse turn-over time $C^{1/2} \epsilon ^{1/3} L^{-3/2} $ for the vortices comparable to the system size $L$. Then the flow coherent over the system size (the condensate) appears \cite{Smith_Yakhot_93,Molenaar04,Chertkov07,Sommeria86,Paret_Tabeling_98,Shats07} with the velocity estimated from the energy balance, $\alpha V^2 \approx 2\epsilon$, which gives  $s \cong V/L_s \cong L^{-1}_s \sqrt{2\epsilon / \alpha}$ and

\begin{equation}\label{Eq:k_t}
 k_t = \pi / l_t \cong \pi L_s^{-3/2} (C \alpha /2)^{-3/4} \epsilon ^{1/4}.
\end{equation}
\noindent Note that this is not the condition that the turnover time at $l_t$ is $\alpha ^{-1}$, as in \cite{Smith_Yakhot_93}; incidentally Eq.~\ref{Eq:k_t} gives a correct estimate ($k_t \cong 1$) for their conditions. The spectrum $E(k) \propto k^{-3}$ at $k<k_t$ is due to the condensate \cite{Smith_Yakhot_93, Shats07, Chertkov07}, while $E(k)=C\epsilon ^{2/3}k^{-5/3}$ is expected at $k_t<k<k_f$.

\begin{figure}
\includegraphics[width=8.5 cm]{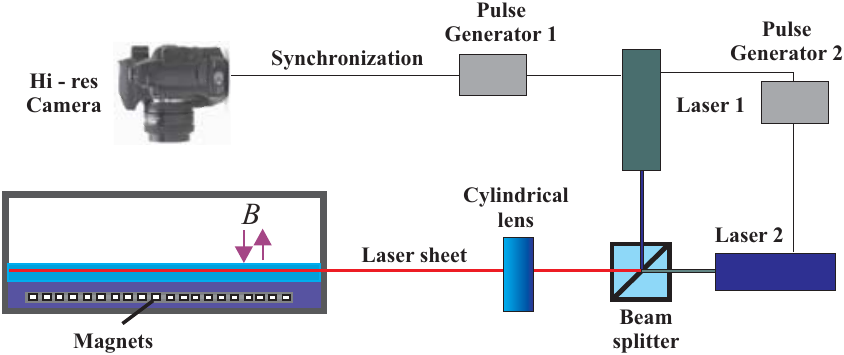}
\caption{\label{fig_1} Experimental setup}
\end{figure}

Here we report the experiment in which the strength and the spectral extent of the condensate are varied by changing either $\alpha$ or $L$. The experimental setup shown in Figure~\ref{fig_1} is similar to those described in \cite{Paret_Tabeling_98, Shats07, Chen06} but has a substantially larger number of forcing vortices (up to 900), higher spatial resolution and larger scale separation ($L/l_f \approx 30$). Turbulence is generated electromagnetically in stratified thin fluid layers whose thicknesses are varied to achieve different $\alpha$. A heavier non-conducting fluid (Fluorinert, specific gravity SG$=1.8$) is placed at the bottom. A lighter conducting fluid, NaCl water solution (SG$=1.03$), is placed on top. Non-uniform magnetic field is produced by a square matrix of $30 \times 30$ permanent magnets ($10\;$mm apart). The electric current flowing through the top (conducting) layer produces $J \times B$-driven vortices which generate turbulence. Square boundaries with $L = (0.09-0.24)\;$m are used. To visualize the flow, imaging particles are suspended in the top layer and are illuminated by a 1 mm laser sheet parallel to the fluid surface. Laser light scattered by the particles is filmed from above using a 12 Mpixel camera. Green and blue lasers are pulsed for 20 ms consecutively with a delay of (20-150) ms. In each camera frame, two laser pulses produce a pair of images (green and blue) for each particle. The frame images are then split into a pair of images according to the colour. The velocity fields are obtained from these pairs of images using particle image velocimetry. The damping rate (in the range $\alpha = 0.05-0.5\;$s$^{-1}$) is estimated from the decay of the total kinetic energy after switching off the forcing: $E\propto e^{-\alpha t}$.

\begin{figure}
\includegraphics[width=6.0 cm]{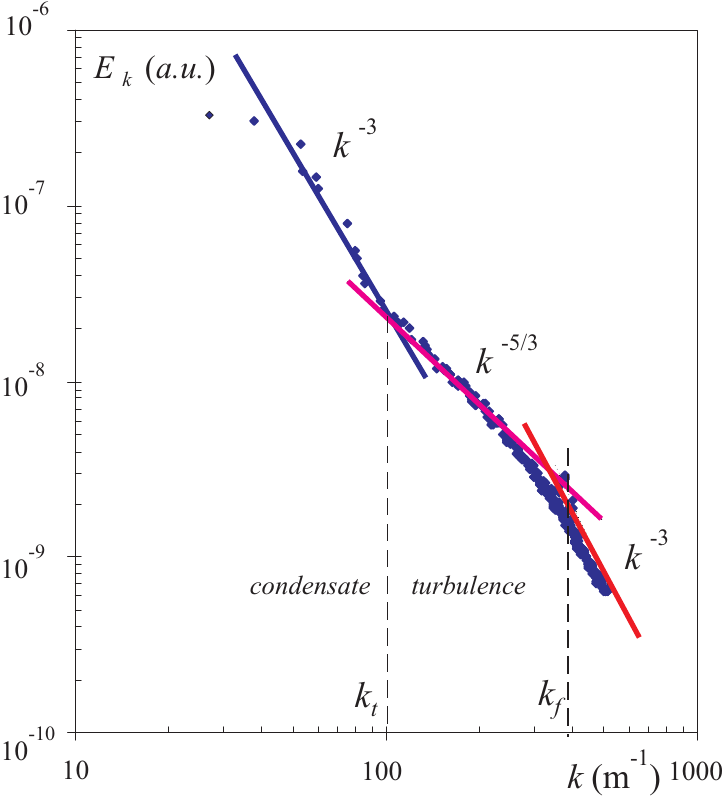}
\caption{\label{fig_2} Kinetic energy spectrum measured for the largest box $L = 0.235$ m and intermediate damping $\alpha = 0.16 s^{-1}$. The guide lines show the power laws for different ranges:  $k^{-3}$ vorticity cascade, $k^{-5/3}$ energy cascade and $k^{-3}$ condensate.}
\end{figure}

\begin{figure}
\includegraphics[width=6.5 cm]{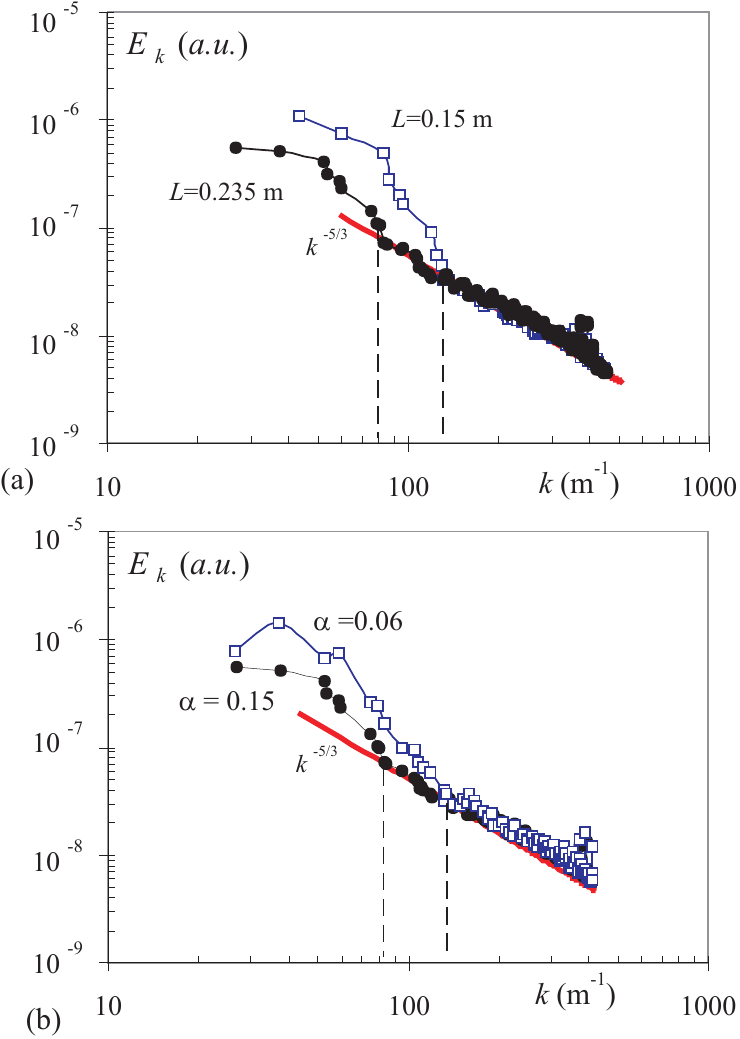}
\caption{\label{fig_3} Kinetic energy spectra (a) for different box sizes at $\alpha = 0.15$ s$^{-1}$, and (b) for different linear damping rates at $L = 0.235$ m.}
\end{figure}

Figure~\ref{fig_2} shows the energy spectrum measured for large $L = 0.235\;m$ and  an intermediate $\alpha=0.16\,s^{-1}$. The force wavenumber is $k_f \approx 400\,$m$^{-1}$; $E(k) \propto k^{-3}$ at $k > k_f$ , while $E(k) \propto k^{-5/3}$ at $k_t < k < k_f$. At $k < k_t \approx 80\;$m$^{-1}$, in the condensate range, the spectrum is steeper and close to $k^{-3}$. Due to the condensate, the spectrum has $k^{-3}$ and $k^{-5/3}$ ranges for the large and intermediate scales respectively, similarly to the Nastrom-Gage spectrum \cite{NG84}. Spectra for different $L$ and $\alpha$ are shown in Fig.~\ref{fig_3}. At fixed $\alpha = 0.15\,$s$^{-1}$, the knee of the spectrum shifts from $k_t \approx 80$ m$^{-1}$ for $L = 0.235$ m to $k_t \approx 135\ $ m$^{-1}$ for $L = 0.15\,$m, Fig.~\ref{fig_3}(a). For fixed $L$, linear damping affects $k_t$ as shown in Fig.~\ref{fig_3}(b). Going from $\alpha = 0.15\ $ s$^{-1}$ to $\alpha = 0.06\ $ s$^{-1}$, changes $k_t$ from $ 80\ $ m$^{-1}$ to $k_t \approx 130\ $m$^{-1}$. These observations are in a good qualitative agreement with (\ref{Eq:k_t}). By further reducing $L$, we achieve a regime when $k_t \approx k_f$, and the $k^{-5/3}$ range disappears, such that the entire spectrum is $E(k) \propto k^{-3}$, both above and below $k_f$, as in \cite{Shats05}. Therefore, we can control the shape of the spectrum and the relative strength of the condensate with respect to  turbulence.

\begin{figure}
\includegraphics[width=8.5 cm]{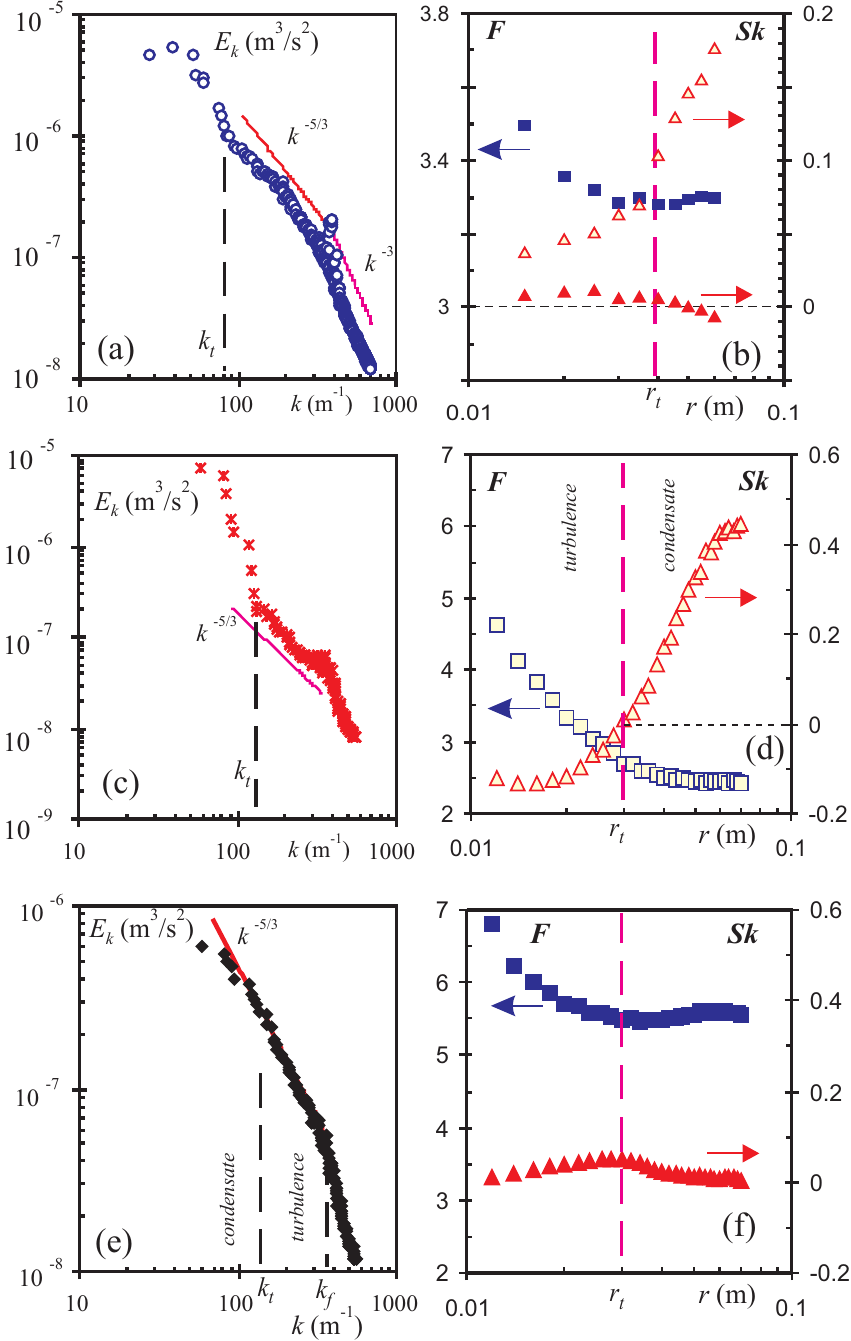}
\caption{\label{fig_4} (a,b) Weak condensate case, (c-f) Strong condensate. (a) and (c) are the kinetic energy spectra. (b) Normalized velocity moments for weak condensate: skewness $Sk$ (triangles) and flatness $F$ (squares). The moments are computed before (open symbols) and after (solid symbols) the mean flow is subtracted from individual velocity fields. (d) Skewness (triangles) and flatness (squares) for the case of strong condensate before subtracting the mean flow. After subtracting mean flow in the stronger condensate case: kinetic energy spectrum (e), (f) skewness (triangles) and flatness (squares).}
\end{figure}

We now analyze two regimes, a weak and a strong condensates, whose spectra are shown in Fig.~\ref{fig_4}(a,c). A weak condensate of Fig.~\ref{fig_4}(a) was generated at $\alpha = 0.3\,s^{-1}$ and $L = 0.235\,m,$  while a stronger condensate of Fig.~\ref{fig_4}(c) was obtained at $\alpha = 0.15\,s^{-1}$ and $L = 0.15\,m$. For the weak condensate case, Fig.~\ref{fig_4}(b) shows normalized velocity moments. Skewness, $Sk=S_{3L}/S_{2L}^{3/2}$ (open triangles) is positive and it is in the range of $Sk=0 \div 0.2$ (skewness of the Gaussian process is $Sk = 0$). Both longitudinal $S_{3L}=\langle(\delta V_L)^3\rangle$ and transverse $S_{3T}=\langle \delta V_L(\delta V_T)^2\rangle$ moments are positive in the entire range of scales, in agreement with expectations for the inverse energy cascade. The $Sk(r)$ dependence has a knee at about $r = 0.04\ $m, which corresponds to the knee in the spectrum at $k_t \approx 80\ $m$^{-1}$, Fig.~\ref{fig_4}(a). The Kolmogorov constant is determined as $C = E(k) \epsilon ^{-2/3} k^{5/3}$, where $\epsilon = S_3/r$. In the weak condensate, at $r < \pi /k_t \approx 0.04\ $m, we have $C=5.6$, which is close to the values $C =5.8 \div 7$ previously obtained in numerical simulations of 2D turbulence (see \cite{Boffetta00} and references therein).  At larger separations, $r > \pi/k_t$, the function $Sk(r)$ grows fast and cannot be approximated by a constant. 

The third-order velocity moment differs markedly in the presence of strong condensate. Fig.~\ref{fig_4}(d) shows that skewness (open triangles) varies in this case in a much wider range, $Sk =-0.2 \div 0.5$. Flatness, $F=S_{4L}/S_{2L}^2$ (open squares), also varies, in the range $F=2.5 \div 5$, while its Gaussian value is $F=3$. For the stronger condensate, the spectrum scales as $E(k) \propto k^{-5/3}$ in the range $k_f > k > k_t \approx 125$ m$^{-1}$. $S_3(r)$ and $Sk$  change sign at $r_t = \pi/k_t$ (Fig.~\ref{fig_4}(d)). Such an $S_3$ dependence on $r$ resembles $S_3(r)$ measured in the lower stratosphere \cite{Cho_Lindborg01}. Note that in our case all the driving comes from small scales and there is no direct cascade at all, yet $S_3$ is strongly modified compared with the weak condensate case. 

The generation of the mean flow can be revealed by a temporal averaging of the instantaneous velocity fields: $\bar{V}(x,y)= \left ( 1/N \right )\sum^{N}_{n=1}V(x,y,t_n)$. The power spectrum of the mean flow \cite{Chertkov07,Shats07} is close to $\bar{E}(k) \propto k^{-3}$. The velocity field contains both the mean component and turbulent velocity fluctuations: $\delta V = \delta \bar{V}+\delta \tilde{V}$. From the data corresponding to Fig.~\ref{fig_4}(c) (strong condensate), we estimate that $\langle(\delta V)^2 \rangle$ differs from $\langle(\delta \tilde{V})^2 \rangle$ by about 20-30\%. However $\langle(\delta V)^3 \rangle$ and $\langle(\delta \tilde{V})^3 \rangle$ differ by orders of magnitude and even the sign can be different. Note that signs, values and functional dependencies $S_3(r)$ vary a lot for different topologies of the condensate flows and also depend on the mean shear in such a flow. 

To recover the statistical moments of the turbulent velocity fluctuations we take $N = 350$ instantaneous velocity fields, subtract their mean flow and then compute the Fourier spectrum and the structure functions. The result for the weak condensate case is shown as solid symbols in Fig.~\ref{fig_4}(b). Mean subtraction brings skewness (solid triangles) much closer to its Gaussian value of $S=0$, while flatness (solid squares) is only slightly higher than its value in isotropic turbulence, $F = 3$  \cite{Paret_Tabeling_98}. 

The result for the strong condensate is shown in Figs.~\ref{fig_4}(e,f). The subtraction of the mean restores the $k^{-5/3}$ range. In fact, the $E(k)$ scatter in Fig.~\ref{fig_4}(e) is less than in the total spectrum of Fig.~\ref{fig_4}(c). The subtraction has even more dramatic effect on the higher moments. As seen in Fig.~\ref{fig_4}(f), there is less variability in both skewness and flatness. $S_3$ is now positive and it is a linear function of $r$ in the "turbulence" range, at $r < r_t$. The spectral energy flux is deduced as $\epsilon = S_3/r$. The value of the Kolmogorov constant $C = E(k) \epsilon ^{-2/3} k^{5/3} \approx 7.7$ appears to be slightly higher than in the weak condensate case, but is still close to the values obtained in numerical simulations. The recovery of the linear positive $S_3(r)$ has also been observed at even stronger condensates. 

We observe that the condensate formation substantially increases flatness: from about the Gaussian value of 3 for the weak condensate (solid squares in Fig.~\ref{fig_4}(b)) to $F\simeq5.5$ for the strong condensate (solid squares in Fig.~\ref{fig_4}(f)). This can be explained as follows: condensate shear suppresses turbulence level by stretching and destroying vortices. Note that strong vortices with the vorticity exceeding the external shear survive. Therefore, strong fluctuations are affected by condensate less than the mean level, which increases $F$.

It is important to note that similarity of our spectra to those of \cite{Lilly83, NG84} does not necessarily mean that $k^{-3}$ spectrum at large scales in the Earth atmosphere is also fed by the inverse cascade. To establish whether this is the case, one needs to analyze the atmospheric data in the way described here: subtract the coherent flow, recalculate the second and the third moment of fluctuations and use Eq.~\ref{Eq:k_t}. It is likely that the baroclinic (large-scale) instabilities play a role in forcing the large-scale flows. To model an external large-scale forcing in our experiments we added a large magnet on top of the small-scale forcing (as described in \cite{Shats07}) and found that the modifications in $S_3$ are similar to those when the large-scale flow is formed via spectral condensation. The mean subtraction recovers the energy flux from small to large scales in both cases. Similarly, the mesoscale turbulence in the Earth atmosphere should be affected by the large-scale flow regardless of its origin.
Let us stress that our experimental system is much simpler than the Earth atmosphere. 
With regard to the meso-scale and large-scale atmospheric motions, there are two most important differences, namely the character of stratification and the absence of rotation in our system. Recent numerical simulations (see \cite{Brethouwer07} and
references therein) show that stratification may enforce a 3D
dynamics and the forward energy cascade. On the other hand, recent
experimental studies of  decaying turbulence suggest a strong role
of rotation in establishing a quasi-2D regime in which geostrophic
dynamics is dominant and the energy cascade is inverse (regime of low Froude and Rossby numbers) \cite{Praud06}. More experiments in forced
turbulence are needed to understand the competing effects
of rotation and stratification along with the complex interplay
between turbulence and waves, resonant wave-wave interactions,
etc. True nature of atmospheric turbulence both at large and meso-scales can only be revealed by the atmospheric measurements.  What we have shown here is the need to separate mean flows and fluctuations to recover the energy flux.

We conclude by stating that the condensate strongly modifies both turbulence level and its statistics; different velocity moments are affected at different scales.

\begin{acknowledgments}
We are grateful to V.V. Lebedev, M. Chertkov and R.E.
Ecke for useful discussions and to D. Byrne for the help with the
data analysis. This work was supported by the Australian Research Council's Discovery Projects funding scheme (DP0881544), Israeli Science Foundation and Minerva Einstein Center, and also in part by the National Science Foundation under Grant No. PHY05-51164.
\end{acknowledgments}

\end{document}